\documentclass[twocolumn,epjc3]{svjour3}
\smartqed  
\usepackage[english]{babel}
\usepackage{epsfig}
\usepackage{wrapfig}
\usepackage{graphicx}
\usepackage{amssymb}
\usepackage{amsfonts}
\usepackage{amsmath}
\usepackage{longtable}
\usepackage{rotating}
\usepackage{lscape}
\usepackage{multirow}
\usepackage{float}
\usepackage{bm}

\newcommand\lam{\lambda}

\journalname{Eur. Phys. J. A}

\begin{document}
\title{Nature of $\bm{S}$-wave $\bm{NN}$ interaction \\ and dibaryon production at nucleonic resonance thresholds}
\author{V.I. Kukulin\thanksref{e1,addr1}
\and
O.A. Rubtsova\thanksref{e2,addr1}
\and
M.N. Platonova\thanksref{e3,addr1}
\and
V.N. Pomerantsev\thanksref{e4,addr1}
\and
H. Clement\thanksref{e5,addr2}
\and
T. Skorodko\thanksref{addr2}
}
\thankstext{e1}{e-mail:kukulin@nucl-th.sinp.msu.ru}
\thankstext{e2}{e-mail:rubtsova-olga@yandex.ru}
\thankstext{e3}{e-mail:platonova@nucl-th.sinp.msu.ru}
\thankstext{e4}{e-mail:pomeran@nucl-th.sinp.msu.ru}
\thankstext{e5}{e-mail:heinz.clement@uni-tuebingen.de}

\institute{
Skobeltsyn Institute of Nuclear Physics, Lomonosov Moscow State
University, Leninskie Gory 1/2, 119991 Moscow, Russia\label{addr1}
\and
Physics Institute and Kepler Center for Astro and Particle Physics,
Eberhard--Karls--University T\"ubingen, Auf der Morgenstelle~14, D-72076
T\"ubingen, Germany \label{addr2} }


\date{Received: \today / Accepted: date}

\maketitle

\begin{abstract}
Phase shifts and inelasticity parameters for $NN$ scattering in
the partial-wave channels ${}^3S_1$--${}^3D_1$ and ${}^1S_0$ at
energies $T_{\rm lab}$ from zero to about 1 GeV are described
within a unified $NN$ potential model assuming the formation of
isoscalar and isovector dibaryon resonances near the $NN^*(1440)$
threshold. Evidence for these near-threshold resonances is
actually found in the recent WASA experiments on single- and
double-pion production in $NN$ collisions. There, the excitation
of the Roper resonance $N^*(1440)$ exhibits a structure in the
energy dependence of the total cross section, which corresponds to
the formation of dibaryon states with $I(J^\pi)=0(1^+)$ and
$1(0^+)$ at the $NN^*(1440)$ threshold. These two $S$-wave
dibaryon resonances may provide a new insight into the nature of the
strong $NN$ interaction at low and intermediate energies.

\end{abstract}

\section{Introduction}  \label{intro}
The traditional point of view on the strong $NN$ interaction at
low energies ($T_{\rm lab} \lesssim 350$~MeV) is based on the
classic Yukawa concept \cite{Yukawa} suggesting $t$-channel meson
exchanges between nucleons. Later on, this idea of Yukawa has been
realized in the so-called realistic $NN$ potentials
\cite{Nijm,Arg,CDB}. Recently the realistic $NN$ potentials (of
the second generation) have been replaced by the Effective Field
Theory (EFT) which can treat single and multiple meson exchanges
more consistently~\cite{Eppelbaum,Machl2}. However when the energy is
rising beyond 350 MeV, the numerous inelastic processes enter the
game and the application of the traditional approach meets many
serious problems. Importantly, most of them are related to our poor
understanding of the short-range $NN$ interaction and the
corresponding short-range two- and many-nucleon correlations in
nuclei and nuclear matter \cite{Baldo}.

From the general point of view, these problems should be tightly
interrelated to the quark structure of nucleons and mesons. On the
other hand, the consistent treatment of the intermediate-energy
$NN$ interaction, especially for inelastic processes, within the
microscopic quark models is associated to so enormous difficulties
\cite{Faess,Yamauchi,Stancu}, that nowadays we have to limit
ourselves with some phenomenological or semi-phenomenological
treatment. However it is still possible to use some hybrid
approach and to combine the meson-exchange treatment for
the long-range $NN$ interaction with the quark-motivated model for the
intermediate- and short-range interaction \cite{Beyer}. Such a
model can be naturally based on the assumption about the six-quark
bag (or dibaryon) formation at sufficiently short $NN$ distances,
where the three-quark cores of two nucleons get overlapped with
each other \cite{PIYAF}. Implementation of this idea does not
require a detailed knowledge of the six-quark dynamics, but only
needs the projection of the six-quark wavefunctions onto the $NN$
channel and an operator coupling the two channels of different
nature, {\it i.e.}, nucleon-nucleon and six-quark ones. So, in the $NN$
channel we can take into account only the peripheral
meson-exchange interaction, while the influence of the internal
$6q$ channel on the $NN$ interaction can be described by a simple
mechanism of an intermediate dibaryon resonance formation with
appropriate $NN \leftrightarrow 6q$ transition form factors
\cite{JPhys2001,KuInt,AnnPhys2010}.

The dibaryon-induced mechanism for the short-range $NN$ interaction
was initially suggested in Ref.~\cite{PIYAF} and quite
successfully  applied to the description of $NN$ elastic
scattering phase shifts and the deuteron properties in
Refs.~\cite{JPhys2001,KuInt}. However, these works did not
consider the inelastic channels and also did not try to identify the
$S$-matrix poles resulting from the fit of the phase shifts using the
dibaryon resonances found experimentally. On the other hand, in
recent years a number of dibaryon resonances have been discovered
which are manifested most clearly in the inelastic
processes~\cite{hcl}.

In Refs.~\cite{FB2019,Yaf2019,Kukulin1}, we elaborated
a unified model that can describe well both elastic phase shifts
and inelasticities in $NN$ scattering in various partial-wave
channels at laboratory energies from zero up to about 1~GeV. Thus,
it has been shown that one can
reproduce quite satisfactorily both elastic and inelastic $NN$
scattering phase shifts in a broad energy range using only a
one-term separable potential with a pole-like energy dependence
(with complex energy) for the main part of interaction and the
one-pion exchange potential (OPEP) for the peripheral part of
interaction. The model was applied to various partial-wave
channels with $L>0$: $^1D_2$, $^3P_0$, $^3P_2$, $^3D_3$--${}^3G_3$
and others, and the theoretical parameters (mass and width) of
dibaryon resonances found from the fit of $NN$ scattering in these
channels turned out to be very close to their experimental values \cite{FB2019,Yaf2019,Kukulin1}.

However, the description of just $S$-wave $NN$ scattering (both
elastic and inelastic) at low and intermediate energies should be
especially sensitive to the assumptions made in the
dibaryon-induced model. In fact, in the case of $S$-wave $NN$
scattering, absence of a centrifugal barrier allows the closest
rapprochement of two nucleons to each other, so the short-range
interaction in the $S$-wave channels provides the strongest impact
to the phase shifts. This is especially true for inelastic
scattering which, in turn, should be governed by the same
mechanism as elastic scattering. Thus, it seems evident that the
quark degrees of freedom should play a major role in the
interaction mechanism. In the present paper, we study in detail
just $S$-wave $NN$ scattering within the dibaryon-induced
approach.

In fact, our model is based mainly on an assumption about the
formation of dibaryon resonances which can be coupled to the
various $NN$ channels. Therefore the existence (or nonexistence)
of such states plays a decisive role in whole our approach. So, it
is worth to briefly discuss the current experimental status of the
dibaryon resonances before we can proceed further with the
theoretical description of $NN$ scattering.

In recent years, many so-called exotic states have been observed
in the charmed and beauty meson and baryon sectors. Common to
these $X$, $Y$, $Z$ and pentaquark states is that they appear as
narrow resonances near particle thresholds constituting weakly
bound systems of presumably molecular character \cite{LHCb}.  A
similar situation is also present in the dibaryonic sector, which
can be investigated by elastic and inelastic $NN$ scattering.

Following the recent observation of the narrow dibaryon resonance
$d^*(2380)$ with $I(J^P) = 0(3^+)$ in two-pion production
\cite{MB,prl2011} and then in $NN$ elastic scattering
\cite{np,npfull}, new measurements and investigations revealed
and/or reconfirmed evidences for a number of states near the
$N\Delta$ threshold. Among these the most pronounced resonance is
the one with $I(J^P) = 1(2^+)$, mass $m \approx$ 2148 MeV and
width $\Gamma \approx$ 126 MeV. Since its mass is close to the
nominal $N\Delta$ threshold of 2.17 GeV and its width is
compatible with that of the $\Delta$ itself, its nature has been
heavily debated in the past, though its pole has been clearly
identified in a combined analysis of $pp$, $\pi d$ scattering and
$pp \leftrightarrow d\pi^+$ reaction \cite{SAID}. For a recent
review about this issue see, {\it e.g.}, Ref.~\cite{hcl}. Very
recently also evidence for a resonance with  mirrored quantum
numbers, {\it i.e.}, $I(J^P) = 2(1^+)$ has been found having a
mass $m =$ 2140(10) MeV and width $\Gamma =$ 110(10) MeV
\cite{D21,D21long}. Remarkably, both these states have been
predicted already in 1964 by Dyson and Xuong \cite{Dyson} based on
$SU(6)$ considerations and more recently calculated in a Faddeev
treatment by Gal and Garcilazo \cite{GG} providing agreement with
the experimental findings both in mass and in width.

Whereas these two states represent weakly bound states relative to
the nominal $N\Delta$ threshold and are of presumably molecular
character with $N$ and $\Delta$ in relative $S$ wave, new evidence
has been presented recently also for two states, where the two
baryons are in relative $P$ wave: a state with $I(J^P) = 1(0^-)$,
$m =$ 2201(5) MeV and $\Gamma =$ 91(12) MeV as well as a state
with $I(J^P) = 1(2^-)$, $m =$ 2197(8) MeV and $\Gamma =$ 130(21)
MeV \cite{ANKE}. The values for the latter state agree with those
obtained before in SAID partial-wave analyses \cite{SAID}. The
masses of these $p$-wave resonances are slightly above the nominal
$N\Delta$ threshold, which is understood as being due to the
additional orbital motion \cite{ANKE}. There is suggestive
evidence for the existence of still further states like a $P$-wave
$I(J^P) = 1(3^-)$ state, for which, however, the experimental
situation is not yet as clear \cite{hcl}. It is also worth
emphasising  that the three resonances $1(2^+)$, $1(2^-)$ and
$1(3^-)$ have been shown to give a sizeable contribution to the
$pp \to d \pi^+$ cross sections and polarisation observables~\cite{Plat_PRD} (the
resonance $1(0^-)$ is not allowed in this reaction by the parity
and momentum conservation).

In the description of $NN$ scattering within the dibaryon-induced
model, we considered first the isovector partial channels $^1D_2$,
$^3P_2$, $^3F_3$ and others, where the  dibaryon resonances near
the $N\Delta$ threshold (respectively, $1(2^+)$, $1(2^-)$,
$1(3^-)$, etc.) can be formed \cite{FB2019,Yaf2019}. We have shown
that these resonances determine almost completely $NN$ scattering
in the respective partial channels at energies from zero to about
600--800 MeV (lab.). Then in the work \cite{Kukulin1} $NN$
scattering in the isoscalar $^3D_3$--$^3G_3$ channels has been
shown to be governed by the $0(3^+)$ dibaryon $d^*(2380)$ which is
located 80 MeV below the $\Delta\Delta$ threshold (and thus can be
treated not as a molecular-like but as a deeply bound
$\Delta\Delta$ state). By analogy, for the $S$-wave partial
channels, with which we are concerned here, the respective
dibaryons could be located near the $NN^*(1440)$ threshold, since
the Roper resonance $N^*(1440)$ has the same quantum numbers as
the nucleon, and an $S$-wave $NN^*$ resonance can easily transform
into an $S$-wave $NN$ state. In comparison to $N\Delta$ dibaryons
which can couple to the isovector $NN$ channels only, both isospin
assignments $I=0$ and $I=1$ are allowed for the $NN^*(1440)$
resonances. So, these resonances, if they exist, can couple to the
$^3S_1$--$^3D_1$ (the deuteron) and $^1S_0$ (the singlet deuteron)
$NN$ channels, respectively.

Fortunately, a strong indication of existence of these two
dibaryon resonances near the threshold of the Roper resonance
excitation have been found in the recent WASA experiments
on single- and double-pion production
in isoscalar and isovector $NN$ collisions~\cite{NNpi,iso}. It
will be demonstrated below that the scenario of dibaryonic resonances
near the $N\Delta$ threshold is not unique, but is repeated at the
$NN^*(1440)$ threshold. And just these resonances determine the
$S$-wave $NN$ scattering at low and intermediate energies.

The paper is organised as follows. In Sec. 2 we briefly outline
the theoretical formalism of the dibaryon-induced model for the $NN$
interaction \cite{FB2019,Yaf2019,Kukulin1} with some modifications
necessary to apply it to $S$-wave scattering. Then in Sec. 3 we
derive the dibaryon parameters from the fit to the phase shifts
and inelasticities in the $^3S_1$--$^3D_1$ and $^1S_0$ channels
and compare them to the experimental data which are discussed in
Sec. 4. We conclude in Sec. 5.


\section{The dibaryon-induced model for the $S$-wave $NN$ interaction}

\begin{figure}[h!]
\centering\epsfig{file=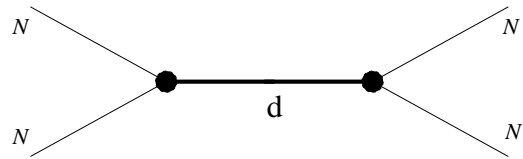,width=0.8\columnwidth}
\caption{\label{nndnn} Diagram illustrating the low-energy $NN$
interaction due to $s$-channel exchange by the deuteron or singlet
deuteron.}
\end{figure}

As is well known, the effective range  approximation for the
low-energy $NN$-scattering leads to the $S$-matrix poles near zero
energy for the triplet $^3S_1$--$^3D_1$ and singlet $^1S_0$
channels. According to the Wigner's idea, one can treat these
$S$-matrix poles as a result of an $s$-channel exchange by the
deuteron or singlet deuteron (see Fig.~\ref{nndnn}).

Then the question arises: whether such $s$-channel  mechanism can
provide description not only of low-energy but also
intermediate-energy $NN$ scattering? The answer is: surely, if
instead of the deuteron pole in Fig.~\ref{nndnn} one will imply a
corresponding dibaryon pole at intermediate energy. The first
attempt to treat the $S$-wave $NN$ scattering at intermediate
energies by the $s$-channel exchange by the dibaryon pole was
undertaken at the beginning of 2000s within the framework of the
dibaryon concept for the nuclear force \cite{JPhys2001}. The
peripheral meson-exchange $NN$ interaction was described via the
so-called external space (or channel) where one deals with
nucleonic and mesonic degrees of freedom. The main short-range
$NN$ attraction is caused by a coupling between the external and
internal channels, where the latter is treated by means of
quark-gluon (or string) degrees of freedom. The rigorous
mathematical formalism to describe such quantum systems combining two
Hilbert spaces (or channels) with completely different degrees of
freedom was developed in the numerous papers of the Leningrad group
\cite{Kuperin}. We refer the reader to Refs.~\cite{JPhys2001,KuInt} where this approach was used to develop a
dibaryon-induced $NN$-interaction model (referred to as a
``dressed bag model'') based on a microscopic six-quark shell
model in a combination with the well-known $^3P_0$ mechanism of
pion production.

\begin{figure}[h!]
\centering\epsfig{file=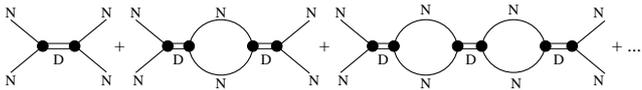,width=\columnwidth}
\caption{\label{diagram} Graphical representation of the $NN$
scattering amplitude driven by the intermediate dibaryon $D$
formation in the $NN$ system.}
\end{figure}

A deeper insight into the structure of the six-quark system in the
internal channel may be gained from the quark-cluster picture
\cite{Nijm6q,ITEP}, where two separated quark clusters, a
tetraquark $4q$ and a diquark $2q$ are connected by a color string
which can vibrate and rotate. In the quark shell-model language,
such a state corresponds to the six-quark configuration
$|s^4p^2[42]_x;L\!=\!0,2;ST\rangle$ with two quarks in the
$p$-shell \cite{PIYAF,JPhys2001}. Being transformed into the
$4q$--$2q$ two-cluster state, it corresponds to the $2\hbar\omega$
excitation of the color string connecting two clusters. So,
coupling between the external and internal channels corresponds to
passing from a bag-like $2\hbar\omega$-excited six-quark state to
$NN$ loops in the external channel (see Fig.~\ref{diagram}). Of course,
the intermediate dibaryon can decay also into inelastic channels (other than $NN$).
In our model, such decays are effectively taken into account through the width $\Gamma_D$
 (see Eq.~(\ref{gamd}) below).
  So that, in Fig.~\ref{diagram},  dibaryon decays into $NN^*$ channel are implicitly
  included in the dibaryon propagator as well.

In Refs.~\cite{FB2019,Yaf2019,Kukulin1}, the dibaryon-induced
model has been generalised further to effectively include the
inelastic processes. Below, for the readers' convenience, we
briefly outline the basic formalism of the dibaryon-induced
model with a special emphasis on $S$-wave $NN$ scattering.
As has been mentioned above, the total Hilbert space of the model
includes the external and internal channels. The external channel
corresponds to the relative motion of two nucleons, while the
internal channel corresponds to the formation of the six-quark (or
dibaryon) state. In the simplest case, the internal space is
one-dimensional, and a single internal state $|\alpha\rangle$ is
associated with the ``bare dibaryon'' having the complex energy
$E_D$. So, the total Hamiltonian has the matrix form:
\begin{equation}
H=\left(
\begin{array}{cc}
h_{NN}& \lam|\Phi\rangle\langle \alpha|\\
\lam |\alpha\rangle \langle \Phi|& E_D|\alpha\rangle \langle
\alpha|\\
\end{array}
\right), \label{Ham}
\end{equation}
where the transition form factor $|\Phi\rangle$ is defined in the
external space and represents a projection of the total $6q$
wavefunction onto the $NN$ channel. In particular, in case
of the coupled spin-triplet $NN$ partial waves $^3S_1$--${}^3D_1$, $|\Phi\rangle$ is a two-component column (see
Ref.~\cite{Kukulin1}).

The external Hamiltonian $h_{NN}$ includes the peripheral
interaction of two nucleons which is given by the one-pion
exchange potential $V_{\rm OPEP}$. Here we use the same form and
the same parameters of $V_{\rm OPEP}$ as in Ref.~\cite{Kukulin1}
\footnote{For the coupled spin-triplet channels
$^3S_1$--${}^3D_1$, we use a bit lower cutoff parameter
$\Lambda_{\pi NN}=0.62$ GeV (instead of $0.65$ GeV employed in
Ref.~\cite{Kukulin1}) which allows for a better fit of the
${}^3D_1$ phase shift.}. For $S$-wave $NN$ scattering, one should
also take into account the six-quark symmetry aspects leading to
an additional repulsive term $V_{\rm orth}$ in the $NN$
potential~\cite{Yaf2019}. Thus, the external Hamiltonian is
represented as a sum of three terms:
 \begin{equation}
 h_{NN}=h_{NN}^{0}+V_{\rm OPEP}+V_{\rm orth},
 \end{equation}
where $h_{NN}^{0}$ is the two-nucleon kinetic energy operator
(which may include the Coulomb interaction for the $pp$ case) and
$V_{\rm orth}$ has a separable form
\begin{equation}
V_{\rm
orth}=\lam_0|\phi_0\rangle\langle \phi_0|. \label{vort}
\end{equation}

The symmetry-induced operator $V_{\rm orth}$ was introduced for
the first time in Ref.~\cite{KK1}. It corresponds to the full or
partial exclusion of the space symmetric six-quark component
$|s^6[6]\rangle$ from the total $NN$ wavefunction and is needed to
fulfil the orthogonality condition between the small
$|s^6[6]\rangle$ and the dominating mixed-symmetry
$|s^4p^2[42]\rangle$ components in the $NN$ system. It has been
shown \cite{KK2} that the operator $V_{\rm orth}$ plays the role
of the traditional $NN$ repulsive core. In fact, this
$s^6$-eliminating potential provides a stationary node in the $NN$
wavefunctions at different energies, and the position of the node
corresponds to the radius  of the repulsive core \cite{YAF13}.

To satisfy the orthogonality condition strictly, one has to take
the limit $\lam_0\to \infty$ in $V_{\rm orth}$. However, since
the $S$-wave $NN$ channels have a strong coupling to the
$S$-wave $NN^*(1440)$ channels near the Roper resonance excitation threshold, the $2\hbar\omega$ excitation in $NN$
relative motion can pass into the $2\hbar\omega$ inner monopole
excitation of the Roper resonance $N^*(1440)$\footnote{In the
quark shell-model language, the $N^*(1440)$ structure corresponds
to the mixture of the $3q$ configurations $0s-(1p)^2$ and $(0s)^2-2s$, both
carrying $2\hbar\omega$ excitation.}. Thus, for such
a strong coupling, there should not be a strict orthogonality
condition for the symmetric configuration $|s^6[6]\rangle$ at
energies near the resonance, and the value of $\lam_0$ should be
finite. There is another good reasoning to this point.
The $S$-wave dibaryon state located
near the  $NN^*(1440)$ threshold can decay into both $NN$ and $NN^*$ channels.
While the relative-motion wavefunction in the $NN$ channel has a stationary node at $r_c=0.5$
 fm similarly to the low-energy $NN$ scattering the $NN^*$ wave function has not got a
 node because the $2\hbar\omega$ excitation in the initial six-quark
  wave function passes into $2\hbar\omega$  inner excitation in the Roper state itself.
Hence, for the channel $NN^*$, the projection operator $V_{\rm orth}$ is not needed.
The especially strong mixing of the $NN$ and $NN^*$ channels happens just in the near-threshold
area where the effect of $V_{\rm orth}$ almost disappears.

So that, we use here the orthogonalising term $V_{\rm orth}$
with the finite values of $\lam_0$. It provides a node in the $NN$ relative motion
wavefunctions at small energies, but at intermediate energies,
$NN$ scattering states have some admixture of the nodeless state
$|\phi_0\rangle$\footnote{In particular, a resonance state in the $NN$
channel may have a noticeable overlap with the state
$|\phi_0\rangle$. The detailed study of this formalism will be
published elsewhere.}. So, in this approach, the finite properly chosen
value of $\lam_0$ provides an effective account of the strong coupling
between the $NN$ and $NN^*(1440)$ channels.

After excluding the internal channel, one gets the effective
Hamiltonian in which the main attraction is given by the energy-dependent pole-like interaction:
\begin{equation}
H_{\rm eff}(E)=h_{NN} +\frac{\lam^2}{E-E_D}|\Phi\rangle \langle
\Phi|. \label{Heff}
\end{equation}
By using the separable form for the energy-dependent part of interaction, one can find explicitly an equation for
the poles of the total $S$-matrix (see details in Refs.~\cite{Yaf2019,Kukulin1}):
\begin{equation}
Z-E_D-J(Z)=0, \label{res_con}
\end{equation}
where the function $J(Z)$ is determined from the matrix element of
the external Hamiltonian resolvent
$g_{NN}(Z)=\left[Z-h_{NN}\right]^{-1}$, {\it i.e.}, $J(Z)=\lam^2\langle
\Phi|g_{NN}(Z)|\Phi\rangle$.

Finally, for the effective account of inelastic processes, we
introduce the imaginary part of the internal pole position
$E_D=E_0-i \Gamma_D/2$, which is energy-dependent and describes
the possible decays of the ``bare''  dibaryon into all inelastic
channels ({\it i.e.}, except for the $NN$ one). For a single decay
channel, the width $\Gamma_D$ can be represented as follows:
\begin{equation}
\Gamma_D(\sqrt{s})=\left\{
\begin{array}{lr}
0,& \sqrt{s}\leq E_{\rm thr};\\\displaystyle
\Gamma_0\frac{F(\sqrt{s})}{F(M_0)},&\sqrt{s}>E_{\rm thr}\\
\end{array}\label{gamd}
\right.,
\end{equation}
where $\sqrt{s}$ is the total invariant energy of the decaying
resonance, $M_0$ is the bare dibaryon mass, $E_{\rm thr}$ is the
threshold energy, and $\Gamma_0$ defines the partial decay width
at $\sqrt{s}=M_0$. For the $S$-wave dibaryon resonance located
near  the $NN^*(1440)$ threshold, the dominant decay channel is
$D\to NN^*(1440)$. Here we take into account only the main decay
mode of the Roper resonance $N^*(1440) \to \pi N$, and thus the
main decay channel of the dibaryon is $D\to \pi NN$. For such a
case, the parametrization of the function $F(\sqrt{s})$ in
Eq.~(\ref{gamd}) has been introduced in Ref.~\cite{Yaf2019}:
\begin{equation}
F(\sqrt{s})=\frac{1}{s}\int_{2m}^{\sqrt{s}-m_{\pi}} \frac{dM_{NN}
q^{2l_\pi+1}k^{2L_{NN}+1}}{(q^2+\Lambda^2)^{l_\pi+1}(k^2+\Lambda^2)^{L_{NN}+1}}.
\label{fpinn}
\end{equation}
Here $\displaystyle
q={\sqrt{(s-m^2_\pi-M^2_{NN})^2-4m_\pi^2M_{NN}^2}}\big/{2\sqrt{s}}$
and $\displaystyle k=\frac{1}{2}\sqrt{M_{NN}^2-4m^2}$ are the pion
momentum in the total center-of-mass frame and the momentum of the
nucleon in the center-of-mass frame of the final $NN$ subsystem with the
invariant mass $M_{NN}$, respectively. In Eq.~(\ref{fpinn}), the
high momentum cutoff parameter $\Lambda$ is used to prevent an
unphysical growth of the width $\Gamma_{\rm inel}$ at high
energies. The possible values of the pion orbital angular momentum
 $l_{\pi}$ with respect to the $NN$ subsystem and the orbital angular momentum of two nucleons $L_{NN}$ are
 restricted by the total
angular momentum and parity conservation. Below, the particular values of
$l_{\pi}$, $L_{NN}$ and $\Lambda$ are adjusted to get the best
fit of inelasticity parameters in the partial $NN$
channels in question.

\section{Results for the $^3S_1$--$^3D_1$ and $^1S_0$ partial-wave channels}

\begin{figure*}[ht] \centering\epsfig{file=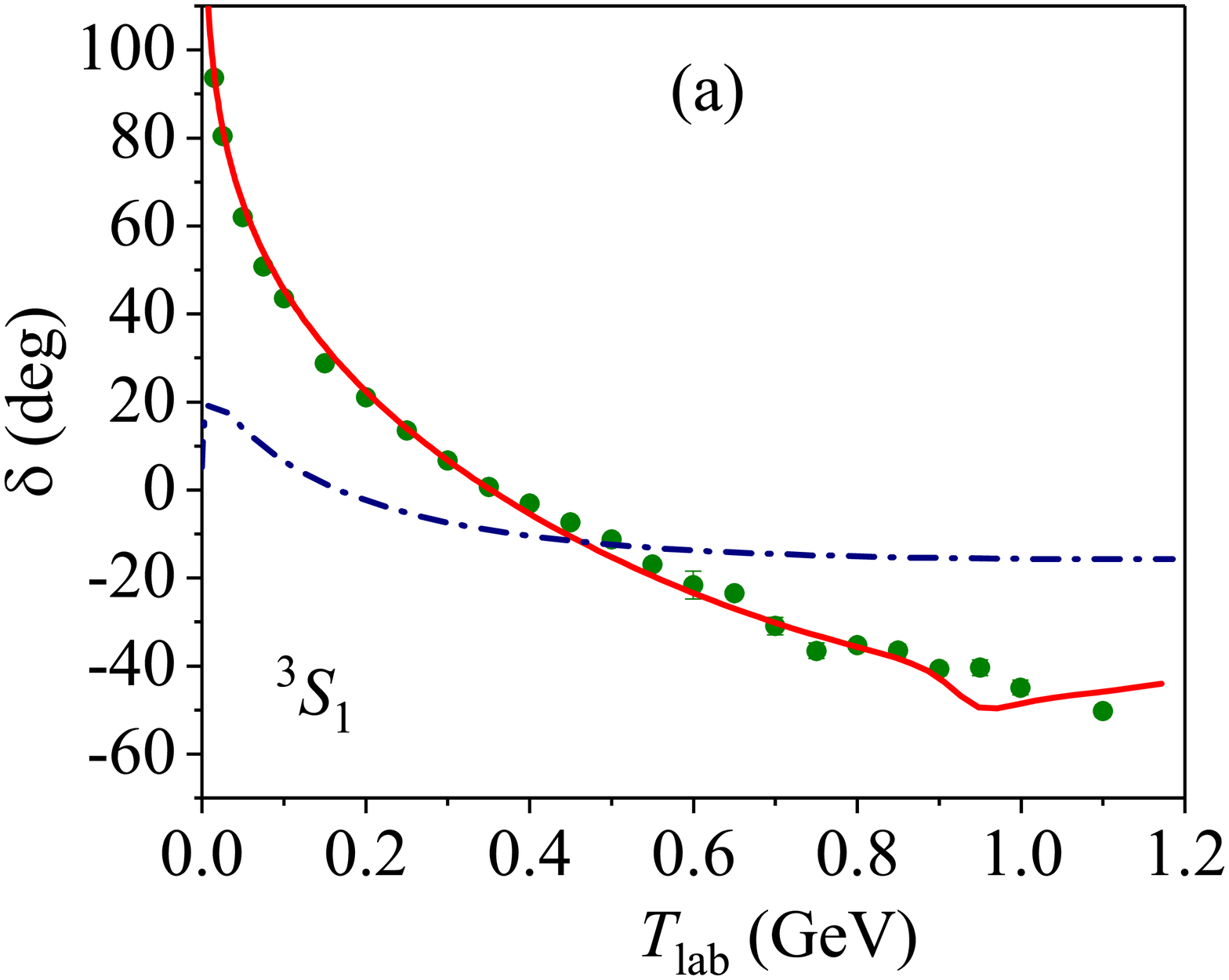,width=0.73\columnwidth}
\hspace{-0.9cm}
\epsfig{file=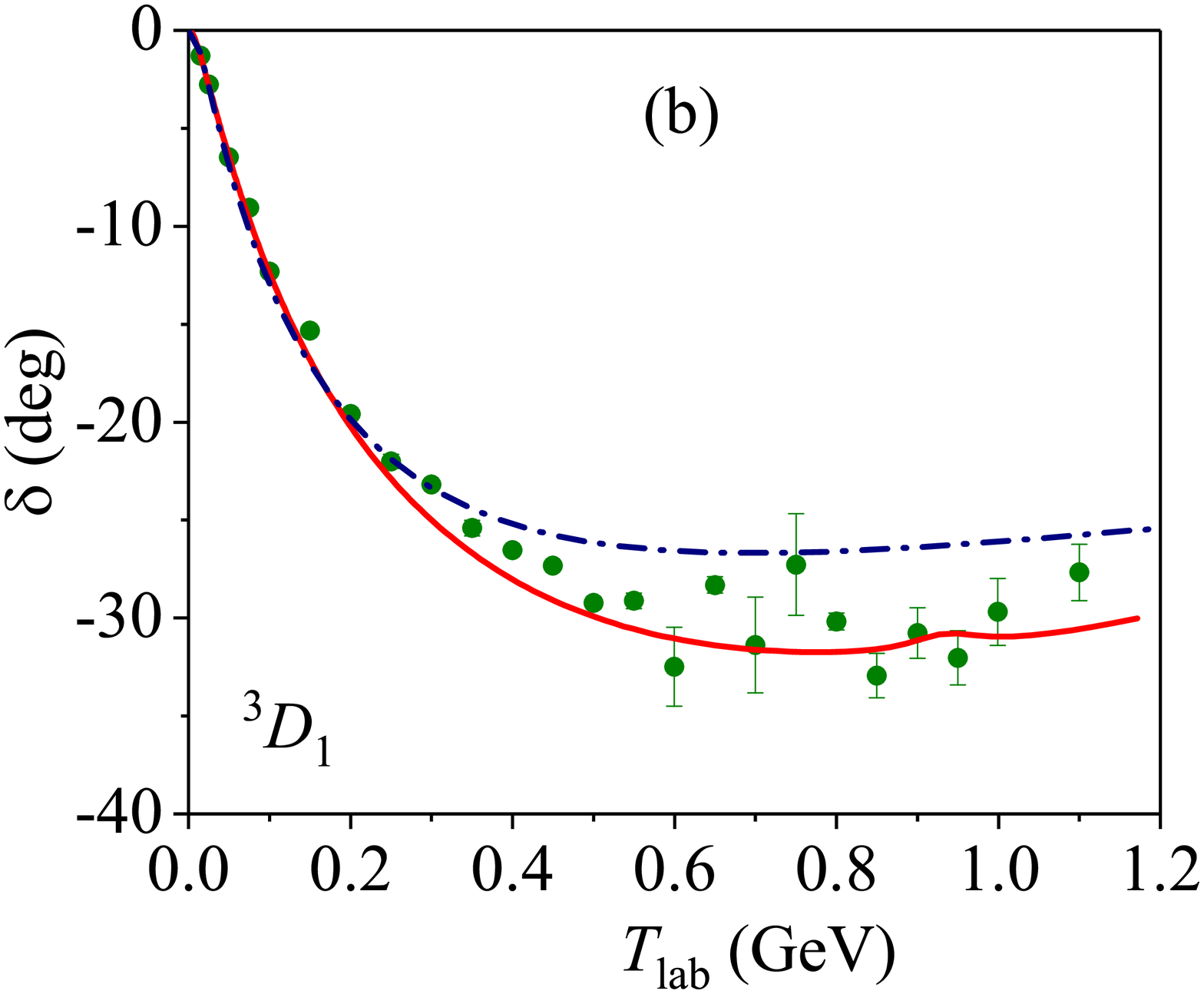,width=0.73\columnwidth}\hspace{-0.9cm}
\epsfig{file=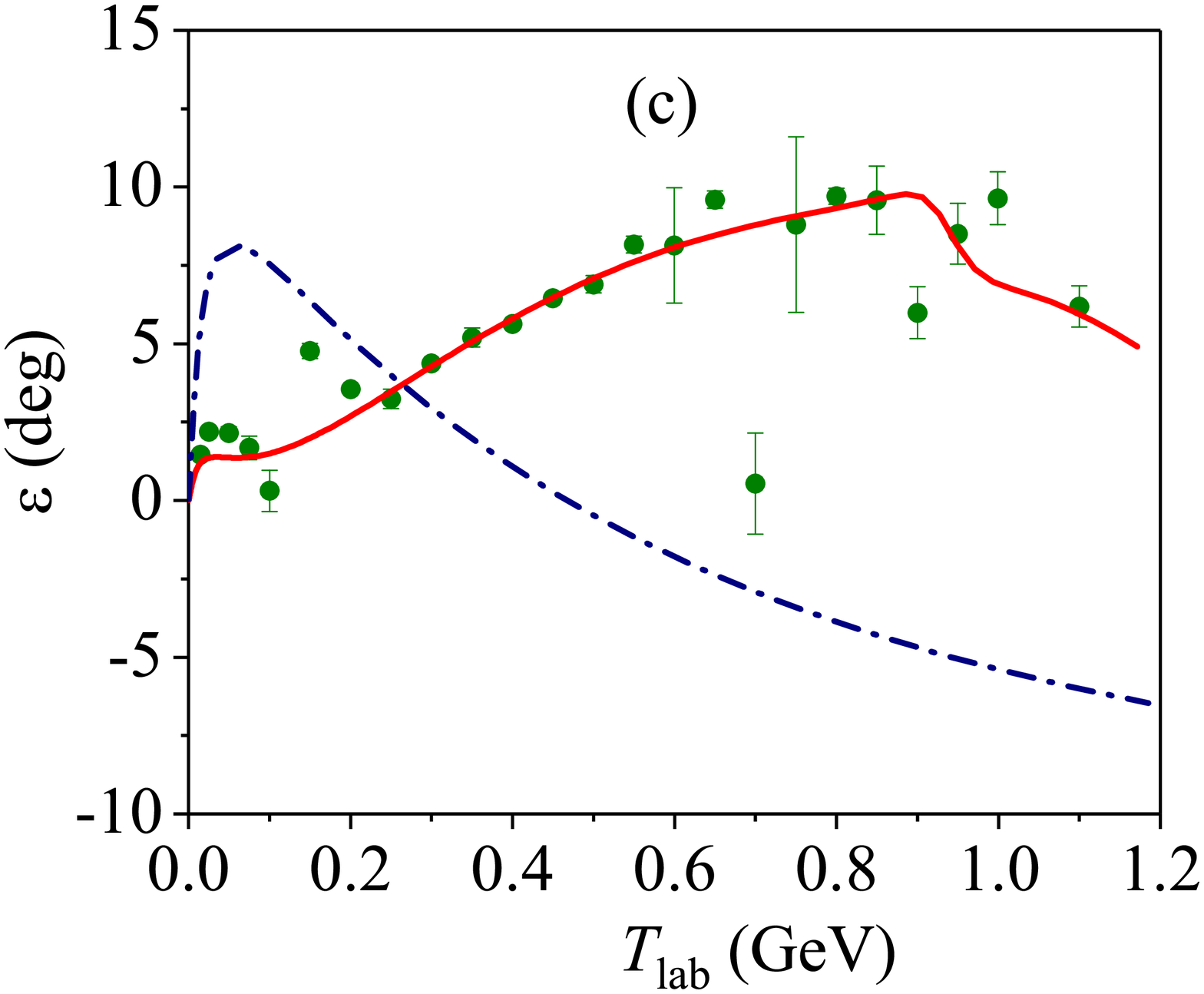,width=0.73\columnwidth}
 \caption{\label{fig3} \small (Color online) Partial phase shifts (a), (b) and
mixing angle (c) for the coupled $NN$ channels $^3S_1$--$^3D_1$
found within the dibaryon model (solid curves) in comparison with
the single-energy SAID PWA~\cite{SAID_PW} (filled circles) and
results for the pure OPEP (dash-dotted curves). }
\end{figure*}
\begin{figure}[h]
\centering \epsfig{file=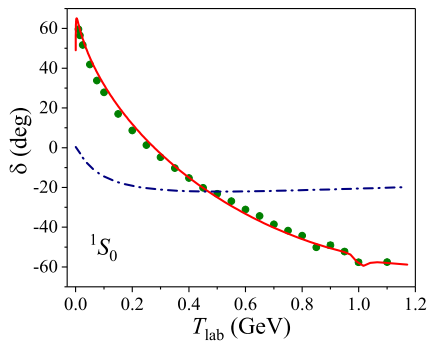,width=0.8\columnwidth}
\caption{\label{fig4} \small (Color online) Partial phase shifts
for the $NN$ channel $^1S_0$ found within the dibaryon model
(solid curves) in comparison with the single-energy SAID
PWA~\cite{SAID_PW} (filled circles) and results for the pure OPEP
(dash-dotted curves).}
\end{figure}
 \begin{figure*}[h]
\centering \epsfig{file=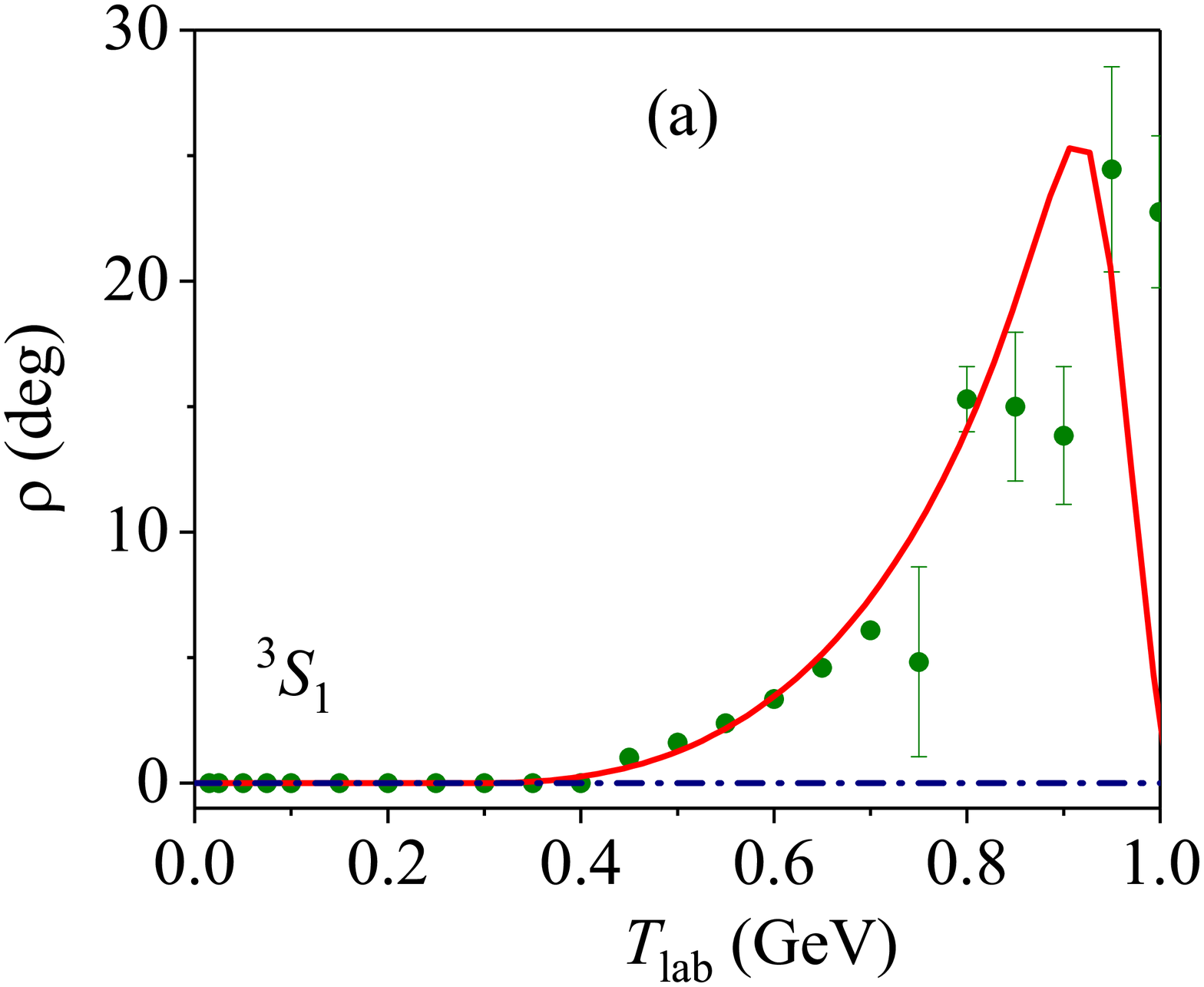,width=0.8\columnwidth}
\epsfig{file=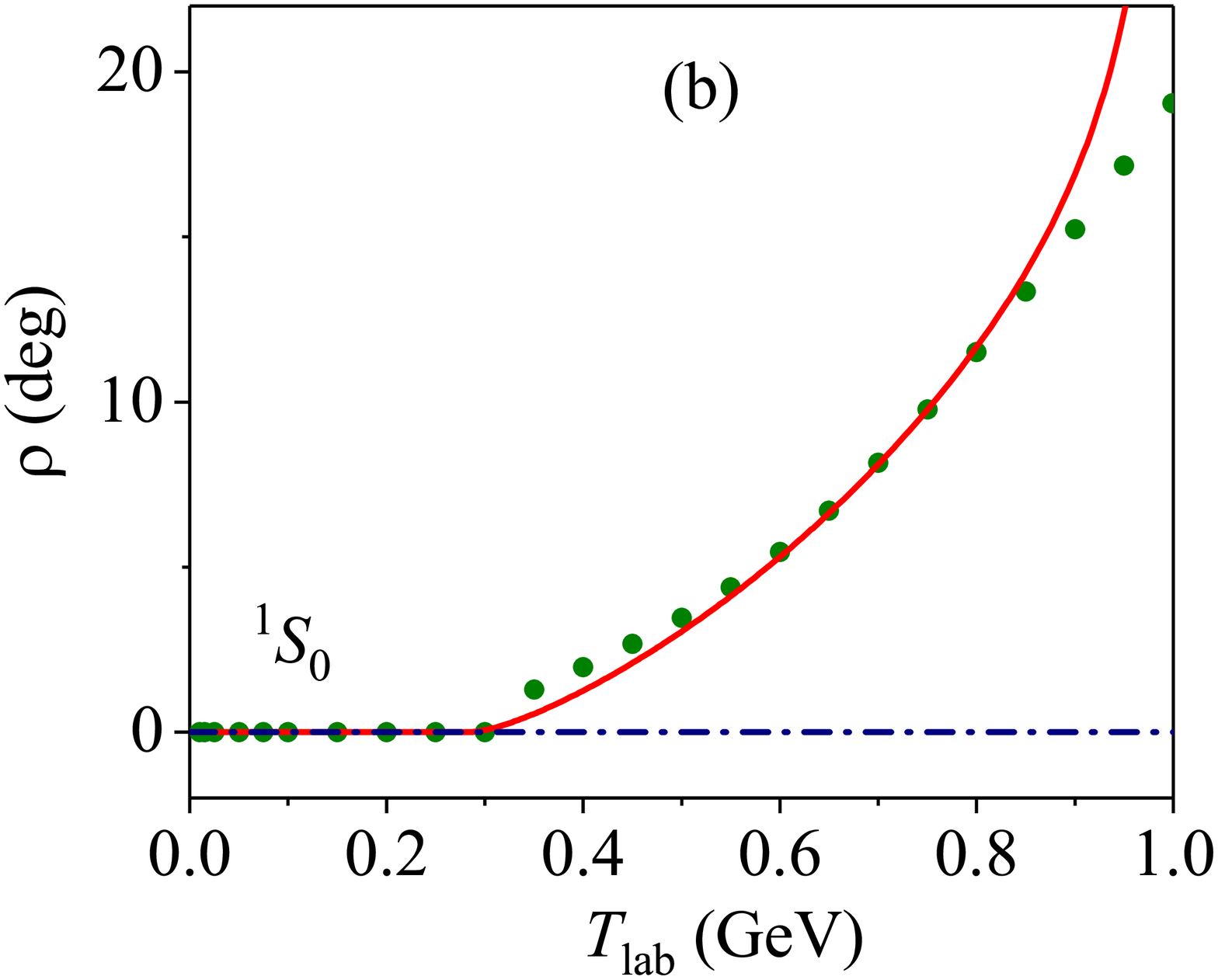,width=0.8\columnwidth}
\caption{\label{fig5} \small (Color online) Inelasticity
parameters for the $NN$ channel $^3S_1$ (a) and $^1S_0$ (b) found
within the dibaryon model (solid curves)  in comparison with the
single-energy SAID PWA~\cite{SAID_PW} (filled circles) and results
for the pure OPEP (dash-dotted curves).}
\end{figure*}

In this section, we present the results of calculations for the $NN$ scattering
phase shifts and inelasticity parameters in the lowest partial-wave channels, viz., $^3S_1$--$^3D_1$ and
$^1S_0$, within the dibaryon-induced model.

For the model form factors entering Eqs.~(\ref{Ham}) and~(\ref{vort}), we have employed the harmonic
oscillator functions with the orbital momentum $L$
and the radial quantum number $n$ equal to the number of nodes in the
$NN$ relative motion wavefunction. In particular, $|\phi_0\rangle$ has the form of the
$S$-wave state with $n=0$ and an effective range $r_0$.
The ``dibaryon'' form factor $|\Phi\rangle$ has two components corresponding to
$S$ and $D$ waves, {\it i.e.}, $|\Phi\rangle=\left(\begin{array}{cc}
\alpha  | \phi_S\rangle\\\beta|\phi_D\rangle\\\end{array}\right)$,
where $\alpha^2+\beta^2=1$. Here $|\phi_S\rangle$ has the same effective range
$r_0$ as $|\phi_0\rangle$, but $n=1$, so, these two
functions are orthogonal to each other. The $D$-wave part of $|\Phi\rangle$ is a nodeless function with the effective
range $r_D$. The potential parameters used for both spin-triplet and spin-singlet partial-wave channels are listed in
Tab.~\ref{Tab1}. Here $\lam_S\equiv \alpha\lam$ and $\lam_D\equiv
\beta\lam$ for the spin-triplet channel. For the dibaryon width defined by Eqs.~(\ref{gamd}) and~(\ref{fpinn}), we used the values $l_\pi=0$, $L_{NN}=1$ and $\Lambda=0.6$ GeV/$c$ for the $^1S_0$ channel and $l_\pi=1$, $L_{NN}=2$ and $\Lambda=1.8$ GeV/$c$
for the coupled $^3S_1$--$^3D_1$ channels. The concrete values of these parameters are important mainly for the fit of inelasticities in the near-threshold region and have a little impact on the overall fit quality.

\begin{table}[h]
\caption{Parameters of the dibaryon model potential for the
lowest spin-triplet and spin-singlet $NN$ partial-wave
channels.\label{Tab1}}
\begin{tabular}{cccccccc}
\hline
&$\lam_0$&$r_0$&$\lam_S$&$\lam_D$&$r_D$&$M_0$&$\Gamma_0$\\
& MeV&fm&MeV&MeV&fm&MeV&MeV\\
\hline
$^3SD_1$& 165 & 0.475 & 248.1 & 65.9 & 0.6 &2275&80\\
$^1S_0$ & 165 & 0.48 & 274.2& - & -& 2300&40\\
\hline
\end{tabular}
\end{table}

The partial phase shifts and mixing angle for the coupled channels
$^3S_1$--$^3D_1$ are shown in Fig.~\ref{fig3} in comparison with
the single-energy (SE) solution of the SAID partial-wave analysis
(PWA) \cite{SAID_PW}. It is seen from Fig.~\ref{fig3} that within the
dibaryon model we can reproduce the PWA data on the
$^3S_1$--$^3D_1$ partial phase shifts and mixing angle in a broad
energy range from zero up to about 1.2 GeV.

The partial phase shifts for the spin-singlet channel $^1S_0$
calculated with the model parameters which are rather close to those used for
the spin-triplet case (see Tab.~\ref{Tab1}) are shown in
Fig.~\ref{fig4} in comparison with the SAID SE data.
Again the dibaryon model allows for the very good description of the partial phase shifts at energies from zero up to 1.2 GeV.

The comparison of inelasticities for the $S$-wave channels with the
SAID single-energy data is presented in Fig.~{\ref{fig5}} (a) and
(b). Here we see reasonable agreement for the $S$-wave
inelasticity parameters with the PWA data up to the energies
corresponding to the resonance position ($T_{\rm lab} \simeq 0.9$
GeV). Thus, the same single-pole model of interaction can
reproduce almost quantitatively both elastic and  inelastic $NN$
scattering in $S$ waves in a broad energy range from
 zero up to about 1 GeV.

In Figs.~{\ref{fig3}}--{\ref{fig5}} the contribution of the pure
OPEP is shown by dash-dotted curves. It is clearly seen that just
the dibaryon  excitation mechanism allows for a reasonable
description of both partial phase shifts and inelasticities for
$S$-wave $NN$ scattering. The coupling with a dibaryon in the
$D$-wave component of the spin-triplet channel $^3S_1$--$^3D_1$ is
weaker, so the dibaryon mechanism makes some important
contribution here only above the inelastic threshold (see
Fig.~{\ref{fig3}} (b)). The situation here is very similar to that
for the $^3D_3$--$^3G_3$ partial-wave channels studied in
Ref.~\cite{Kukulin1}.

It is extremely interesting that the $S$-matrices for the model
$NN$ potentials in both singlet and triplet partial channels have
two poles. For the $^3S_1$--$^3D_1$ case, the first pole
corresponds to the bound state, {\it i.e.}, the deuteron, which is
reproduced rather accurately. The second pole here corresponds to
the dibaryon resonance with the parameters:
\begin{equation}
\label{res_sd}
M_{\rm th }({^3SD_1}) = 2310 \quad {\rm MeV}, \quad \Gamma_{\rm th} ({^3SD_1}) = 157 \quad {\rm MeV}.
\end{equation}

For the $^1S_0$ channel, the first pole is the well-known
singlet deuteron state, while the position of the second one is:
\begin{equation}
\label{res_s}
M_{\rm th }({^1S_0}) = 2330 \quad {\rm MeV}, \quad \Gamma_{\rm th} ({^1S_0}) = 51\quad {\rm MeV}.
\end{equation}

Both these resonance positions are rather close to the
$NN^*(1440)$ threshold\footnote{The difference between the
resonance parameters found here for the $^1S_0$ channel from the
preliminary ones obtained in Ref.~\cite{Yaf2019} is due to the use
of the finite $\lam_0$ in the orthogonalising potential $V_{\rm
orth}$.}. As will be shown below, the
resonance parameters given in Eqs.~(\ref{res_sd}) and (\ref{res_s})
also turn out to be very close to the values derived
from the recent single- and double-pion production experiments (see Sec. 4).
However, the inaccuracy in description of inelasticity parameters
at energies above the resonance position in the considered $NN$
partial-wave channels as well as a too narrow width of the dibaryon
resonance in the $^1S_0$ channel show that a more detailed treatment
of inelastic processes is required within the dibaryon model.

\begin{figure*}[h!]
\centering \epsfig{file=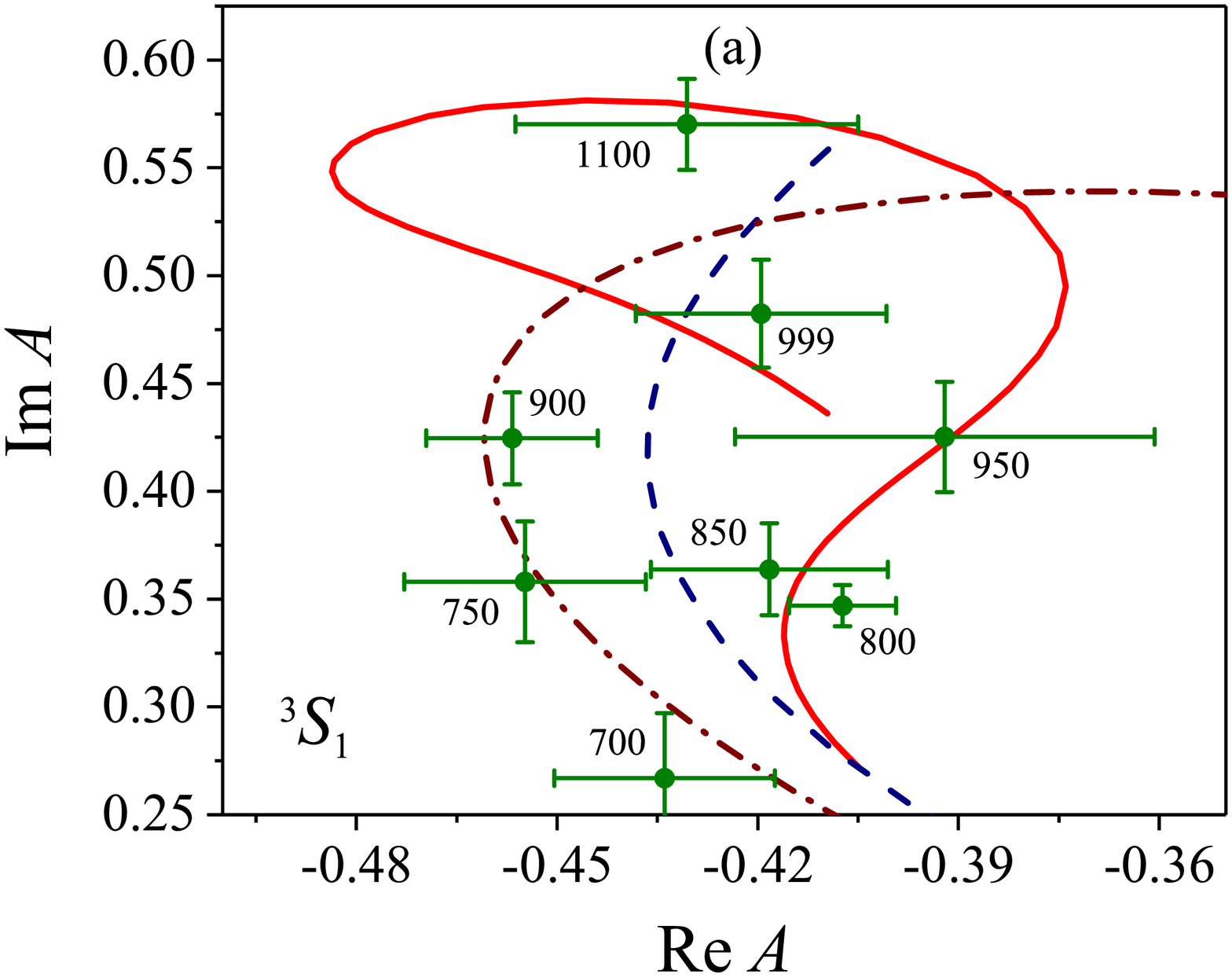,width=0.8\columnwidth}
\epsfig{file=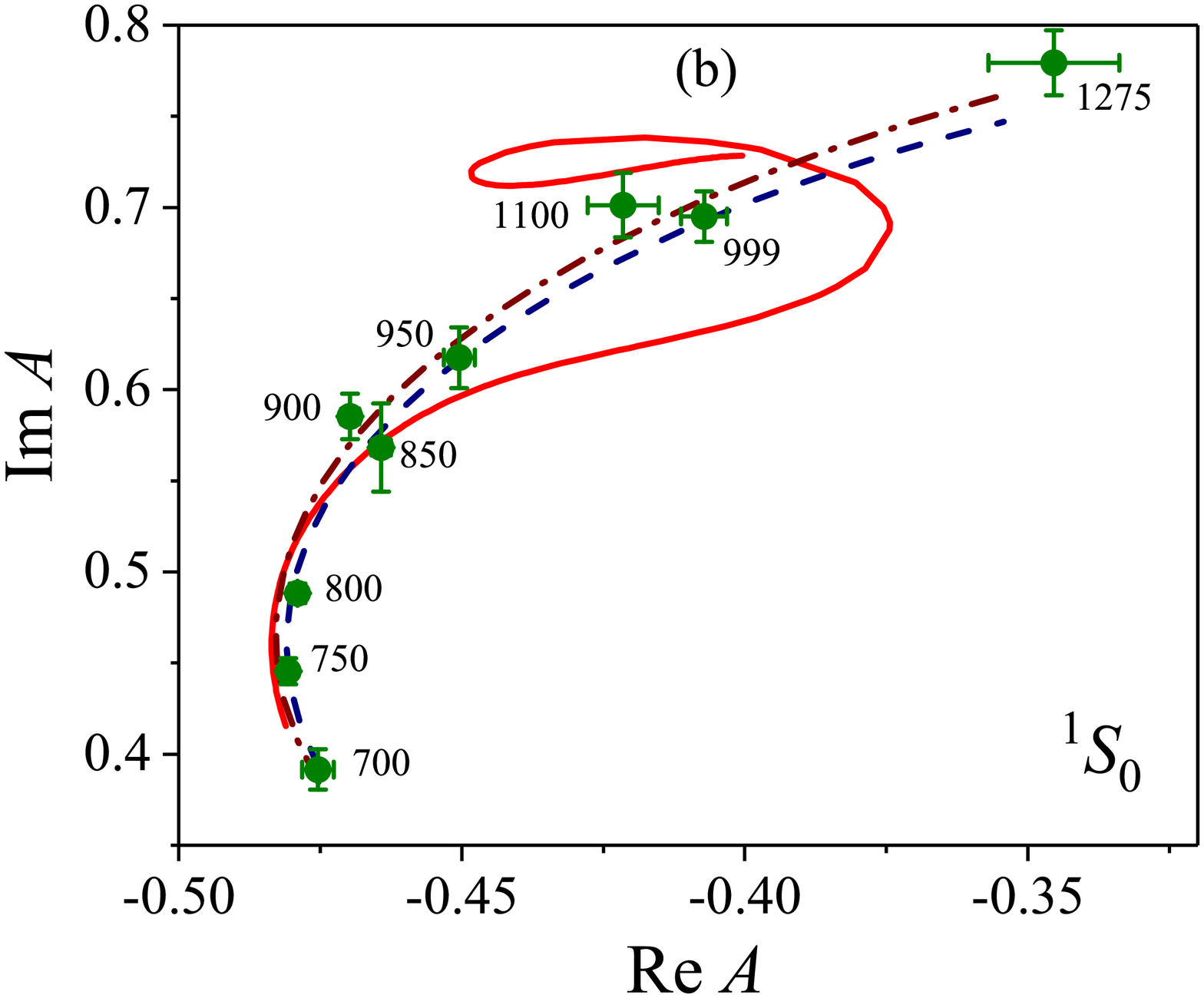,width=0.8\columnwidth}
\caption{\label{Argand} \small (Color online) Argand diagrams for
the $NN$ channels $^3S_1$ (a) and $^1S_0$ (b) found within the
dibaryon model (solid curves)  in comparison with different
solutions of the SAID PWA~\cite{SAID_PW}: single-energy (filled
circles), SM16 (dashed curves) and AD14 (dash-dotted curves). The
numbers near the single-energy points reflect the corresponding
values of the lab. energy  $T_{\rm lab}$ in MeV. }
\end{figure*}

As $s$-channel resonances, the two predicted dibaryon  states have
to display a counter clockwise looping in the Argand diagrams of
amplitudes in $^1S_0$ and $^3S_1$ partial waves. For these partial
waves, two $S$-wave trajectories\footnote{Here the partial
amplitude $A$ is defined as $A=({S_L-1})/{2i}$, where $S_L$ is the
$S$-matrix for the given orbital angular momentum $L$.} are shown
by the solid lines in Fig.~\ref{Argand} in comparison with the
different SAID PWA solutions \cite{SAID_PW}. In fact, we observe
the counter clockwise loopings for the amplitudes found within the
dibaryon model indicating the resonance presence in both cases.

 Since these resonances are highly inelastic, the resonance
loops are rather tiny. The theoretical predictions should be
compared with three PWA solutions of the SAID group, viz., the
single-energy as well as the global solutions AD14 and
SM16~\cite{SAID_PW}. The scatter within the single-energy data as
well as the differences among the various SAID solutions may serve
as an indication for inherent ambiguities in the partial-wave
analysis, especially for the $^3S_1$ channel.  In fact, the
differences between two recent solutions SM16 and AD14 are of the
same order as those differences between theoretical loops and each
of the above SAID solutions. Hence, the absence of the loops in
the current SAID solutions cannot argue against the suggested
dibaryon resonances.

We note that the situation here is much different from that for
the dibaryon resonance d*(2380). In case of the latter, the
partial waves $^3D_3$ and $^3G_3$ were involved, which both carry
large orbital angular momentum and hence have a large impact on
the analyzing power. Since this observable is the only one, which
solely consists of interference terms, it is predestinated to
exhibit substantial effects even from tiny resonance admixtures in
partial waves. Unfortunately, we deal here with $S$-wave
resonances, which make no contribution to the analyzing power due
to the missing orbital angular momentum. Hence this key observable
for revealing loops and resonances is not working here. The only
way out of this dilemma is to look into reactions, where these
highly inelastic resonances decay to, namely single- and
double-pion production. We discuss these processes in the next
Section.

\section{The Roper excitation in $NN$ induced single-
 and double-pion production and the near-threshold dibaryon resonances}

The Roper resonance $N^*(1440)$ excitation appears usually quite
hidden in the observables and in most cases can be extracted from the data
only by sophisticated analysis tools like partial-wave decomposition. By
contrast, it can be observed free of background in $NN$-induced isoscalar
single-pion production, where the overwhelming isovector $\Delta$ excitation
is filtered out by isospin selection as demonstrated by recent WASA-at-COSY
results \cite{NNpi} for the $NN \to [NN\pi]_{I=0}$ reaction. Though the
primary aim of this experiment was the search for a decay $d^*(2380) \to
[NN\pi]_{I=0}$, it also covers the region of the Roper excitation, which is
discussed here.

Since the $\Delta$ excitation is filtered out by the isospin condition, there
is only a single pronounced structure left in the isoscalar nucleon-pion
invariant mass spectrum as seen in Fig.~6 of Ref. \cite{NNpi}, which peaks
at $m \approx$ 1370 MeV revealing a width of $\Gamma \approx$ 150 MeV.
These values are compatible with the pole values for the Roper resonance
deduced in diverse
$\pi N$ and $\gamma N$ studies \cite{PDG}.
Our values for the Roper peak also are in good
agreement with earlier findings from hadronic $J/\Psi \to \bar{N} N\pi$ decay
\cite{BES} and $\alpha N$ scattering \cite{Morsch1,Morsch2}.

\begin{figure}[h]
\centering
\includegraphics[width=0.8\columnwidth]{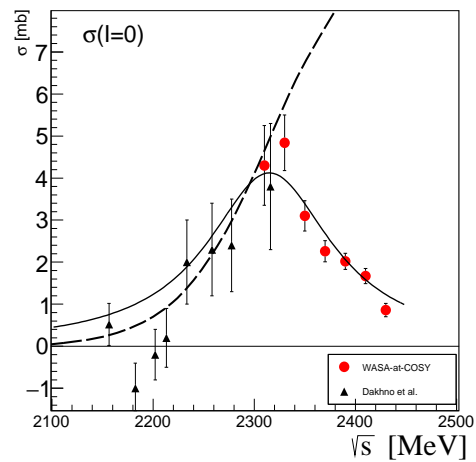}
\caption{\small (Color online)
The $NN$-induced isoscalar single-pion production cross section in dependence
of the total c.m. energy $\sqrt s$. Shown are the recent results from
WASA-at-COSY \cite{NNpi} (solid circles) together with earlier results
\cite{Dakhno} at lower energies. The dashed line shows the expected energy
dependence based on $t$-channel Roper  excitation \cite{NNpi,Luis}, the solid
line a Lorentzian fit to the data with $m =$ 2315 MeV and $\Gamma =$ 150 MeV.
}
\label{fig1}
\end{figure}

The energy excitation function of the measured $NN$-induced isoscalar
single-pion production cross section is displayed in Fig.~\ref{fig1}.
Near threshold the Roper resonance is produced in $S$ wave in relation to the
other nucleon, whereas the pion from the Roper decay is emitted in relative
$p$ wave. Hence we expect for the energy dependence of the total cross section in the
isoscalar $NN \to [NN\pi]_{I=0}$ channel a threshold behavior like for
pion $p$ waves --- as is actually born out by the explicit calculations for the $t$-channel Roper excitation in
the framework of the modified Valencia model \cite{NNpi,Luis}. These
calculations are displayed in Fig.~\ref{fig1} by the dashed
line, which is arbitrarily adjusted in height to the data. The data \cite{NNpi,Dakhno} presented in Fig.~\ref{fig1} follow this expectation
by exhibiting an increasing cross section with increasing energy up to about $\sqrt s \approx$
2.30 GeV. Beyond that, however, the data fall in cross section in sharp
contrast to the expectation for a $t$-channel production process. The observed
behavior is rather  in agreement with a $s$-channel resonance process as
expected for the formation of a dibaryonic state at the $NN^*$ threshold. Due
to the relative $S$ wave between $N$ and $N^*$ as well as due to the isoscalar
nature of this system, it must have the unique quantum numbers $I(J^P) =
0(1^+)$. From a fit of a simple Lorentzian to the data we obtain $m =$
2315(10) MeV and $\Gamma =$~150(30) MeV. The
large uncertainty on the latter results from the large uncertainties of the
data at lower energies (for a fit with a Gaussian, which leads to a width of
170 MeV, see Ref.~\cite{NNpi}). For a more detailed treatment of the resonance
structure one would need to use a momentum-dependent width, which takes into
account the nearby pion production threshold and lowers the resonance cross
section at the low-energy side.

\begin{figure}[h]
\centering
\includegraphics[width=0.8\columnwidth]{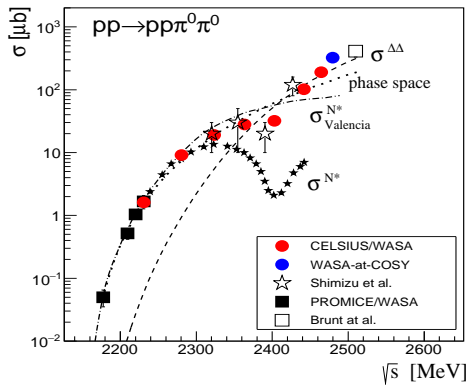}
\caption{\small (Color online) Energy dependence of the total $pp
\to pp\pi^0\pi^0$ cross section. Shown are the data from
CELSIUS/WASA \cite{iso} as well as WASA-at-COSY \cite{tt} (filled
circles),  PROMICE/WASA \cite{JJ} (filled squares), and earlier
work \cite{shim,brunt} (open symbols). The dotted and dash-dotted
lines show the expected energy dependence of simple phase-space
and modelled Roper excitation \cite{Luis}, respectively. The
dashed line shows the $t$-channel $\Delta\Delta$ excitation
\cite{Luis,deldel}, whereas the filled stars display the result of
the isospin decomposition for $N^*$ excitations \cite{iso}. Here,
the first structure is due to the Roper $N^*(1440)$ excitation.
The rerise at higher energies signals higher-lying $N^*$
excitations. } \label{fig2}
\end{figure}

A very similar situation is also observed in $NN$-induced two-pion
production. The situation is particularly clear in the $pp \to
pp\pi^0\pi^0$ reaction, the total cross section of which is
plotted in Fig.~\ref{fig2}. Since ordinary single-$\Delta$
excitation is excluded here due to the necessary production of two
pions, the Roper excitation is the only resonance process at low
energies. Hence we would again expect a phase-space-like (dotted
line) growth of the cross section, which is also born out by
detailed model calculations (dash-dotted line) \cite{Luis,Zou}.
 But the data follow this trend up to $T_p \approx$ 0.9 GeV $(\sqrt s
\approx$ 2.3 GeV). Then the data level off before they increase
again, when the next higher-energetic process, the $t$-channel
$\Delta\Delta$ process with double-$p$-wave emission (dashed line)
starts. Isospin decomposition of the data in the various
$NN\pi\pi$ channels tells us that the energy dependence of the
Roper excitation is experimentally given by the filled star
symbols in Fig.~\ref{fig2} \cite{iso}. Again we see a
resonance-like energy dependence, which indicates a $NN^*(1440)$
molecular system also in this case, but now with quantum
numbers $I(J^P) = 1(0^+)$, $m \approx$ 2320 MeV and $\Gamma
\approx$ 150 MeV. We note that the fading away of the Roper
excitation at energies beyond $T_p \approx$ 0.9 GeV is also in
agreement with the analysis of the corresponding differential
cross sections \cite{deldel,ts}.

\section{Conclusions}

We have shown within the dibaryon-induced model for $NN$ scattering that the $NN$ interaction in the basic spin-singlet
and spin-triplet $S$-wave partial channels at energies $T_{\rm lab}$ from
zero up to about 1 GeV is governed
by the formation of the $I(J^{\pi})=0(1^+)$ and $1(0^+)$ dibaryon
resonances near the $NN^*(1440)$ threshold. This work continues
a series of the previously published papers
\cite{FB2019,Yaf2019,Kukulin1} where the $NN$ interaction in higher
partial waves was shown to be dominated by the intermediate
dibaryon excitation (supplemented by the peripheral one-pion
exchange) in the respective partial channels near the $N\Delta$ or
$\Delta\Delta$ thresholds.

From the energy dependence of $NN$-induced isoscalar single-pion
and isovector double-pion production we see also that both
isospin-spin combinations in the $NN^*(1440)$ system lead
obviously to dibaryonic threshold states at the Roper excitation
threshold --- analogous to the situation at the $\Delta$
threshold. However, compared to the situation there the Roper
excitation cross sections discussed here are small.  Since these
structures decay mainly into inelastic channels, their partial
decay width into the elastic ($NN$) channel should be only a small
fraction of the total width, similarly to the respective branching
ratio for the $N\Delta$ near-threshold states~\cite{hcl}. Despite
this fact, our results show that the contributions of these
dibaryon states to the low- and intermediate-energy $NN$ elastic
scattering are dominating.

On the other hand, at energies below the inelastic thresholds
$NN\pi$ and $NN\pi\pi$, decay of the dibaryons into these channels
is forbidden. However, a strong coupling between the $NN$ and the
closed (virtual)  channels like $N\Delta$, $NN^*(1440)$, $NN\pi$
and $NN\pi\pi$ is still possible. So, at low energies ($T_{\rm
lab} \lesssim 350$~MeV) the coupling of the dibaryon to these
closed channels appears to be strong and thus the whole picture of the
$NN$ interaction at these energies is dominated just by this
coupling. This explains how the intermediate dibaryon formation
near the nucleonic resonance threshold can be the leading
mechanism for the $NN$ interaction at low energies. When the collision
energy is rising and the inelastic channels open, the same
intermediate dibaryons provide single- and double-pion production.
Thus, in the dibaryon-induced approach to the $NN$ interaction, the
elastic and inelastic $NN$ collision processes have a common
origin and can be described via a common mechanism. These results
may provide a novel insight into the nature of the $NN$ interaction at low
and intermediate energies and should be confirmed by further
experimental and theoretical research.

{\bf Acknowledgments.} We are indebted to L. Alvarez-Ruso for
using his code and to I.T. Obukhovsky for fruitful discussions of
the microscopic quark model. The work has been supported by DFG
(grants CL 214/3-2 and 3-3) and the Russian Foundation for Basic
Research, grants Nos. 19-02-00011 and 19-02-00014. M.N.P. also
appreciates support from the Foundation for the Advancement of
Theoretical Physics and Mathematics ``BASIS''.

\end{document}